\documentstyle [12pt] {article}

\catcode`@=12
\begin{document}
\vspace{0.5in}
\oddsidemargin -.375in  
\newcount\sectionnumber 
\sectionnumber=0 
\def\be{\begin{equation}} 
\def\ee{\end{equation}}
\def\ber{\begin{eqnarray}} 
\def\eer{\end{eqnarray}}
\def\ba{\begin{array}} 
\def\ea{\end{array}}
\thispagestyle{empty}
\begin{flushright} October 11, 2000\
\end{flushright}
\vspace {.5in}
\begin{center}
{\Large \bf {Probing For New Physics
            In J/$\Psi$ Decays }\\}
\vspace {.5in}
{\bf  {Xinmin ZHANG
 }\\}
\vskip .3in

%{\it $^2$CCAST (World Laboratory), Beijing 100080 and   \\}
{\it Institute of High Energy Physics (IHEP), Academia Sinica \\}
{\it Beijing 100039, P.R. China\\}

%\vspace{.1in}
\vskip .8in
\end{center}  

\begin{abstract}

Beijing Spectrometer (BES) at 
Beijing Electron Positron Collider (BEPC), IHEP
has
accumulated 2.5 $\times 10^7$  J/$\Psi$ and plans to
increase the number to $10^8 \sim 10^9$ in the near future. In this paper I
review and summarize the recent studies on the possibilities
of probing
for new physics at BES. This includes
the theoretical works on and experimental attempt performed at BES in
searching for flavor changing neutral current, CP violation and lepton
flavor
violation in the processes of J/$\Psi$ decays.

\end {abstract}
\newpage
\baselineskip 24pt
%\section{\bf Introduction}

\section{\bf {Lepton flavor violating J/$\Psi$ decays}}
The J/$\Psi$ rare decay processes
$J/ \Psi ~ \rightarrow {\bar l} l^\prime$
  ( $l, ~ {l^\prime}$ =
$\tau ,~ \mu , e$ ) conserve total lepton number, but violate the
individual
lepton numbers. In the standard model
the lepton flavor symmetries 
 are conserved, but speculated
to be violated in many extensions of the standard model, such as 
supersymmetric standard models, left-right symmetric models and models
where electroweak symmetry is broken dynamically. 
Recent Super-Kamiokande experiment results indicate that neutrinos have 
non-vanishing masses, mix with each others and consequently that lepton
flavor symmetry and/or lepton
number symmetry are broken symmetries. In cosmology mystery of
matter and antimatter asymmetry might be understood in terms of
the brokendown of the lepton number
symmetry together with
the non-perturbative effects (Sphaleron) of the standard electroweak
theory. 

There has been a lot of studies both theoretically and experimentally on
testing the lepton flavor conservation law. At present we have 
various bounds listed in the particle data book from 
$\mu$ decays, $\tau$ and Z gauge boson decays[1]. With a large
sample of J/$\Psi$, BES will be able to make an additional experimental
searching
for lepton flavor violation.

To estimate the branch ratio of lepton flavor violating J/$\Psi$ decays
allowed by the current experimental data, Peccei, Wang and I[2]
took a model-independent approach to new physics and introduced
a four-fermion contact interaction
\begin{eqnarray}
 \frac{4 \pi}{\Lambda^2} {\bar c} \gamma^\mu c
~ {\bar l} \gamma_\mu l^\prime , \ 
\end{eqnarray}
where $\Lambda$ is the new physics cutoff. This effective 
operator is forbidden in the standard model, however will be generated in
theories where lepton flavor is not conserved, such as the 
minimal supersymmetric standard model
with and/or without R parity, models with large extra dimension[3].
Therefore, any observed signal is a direct evidence for non-
standard physics and will
improve our understanding of flavor dynamics, especially
in the lepton sector.

There is no
direct experimental limit on $\Lambda$ in (1). However, 
at one-loop, attaching the neutral gauge boson $Z$ to the charm quark loop
generates an effective coupling of $Z$ to ${\bar l}
l^\prime$ .
%\footnote{Because of the structure of the $\gamma^\mu$ matrix 
%in (1), there is no
%contribution to $\mu \rightarrow e \gamma$ and
%$\tau \rightarrow \mu ({\rm or} e) \gamma$.}. To
%evaluate the rate of $Z \rightarrow {\bar l} l^\prime$,
% we follow the approach to the loop
%calculation in effective theory[1] and
% use the dimensional regularization.  
From the
limits given in the particle data book
on $Br(Z \rightarrow {\bar l} l^\prime )$[1], 
we obtained the lower
bounds on the branch ratio of the $J/ \Psi$
decay into leptons: 

\begin{eqnarray}
&Br(J/\Psi \rightarrow \tau^+ e^- ) <2.7 \times 10^{-5};\\
&Br(J/\Psi \rightarrow \tau^+ \mu^-) <4.9 \times 10^{-5};\\
&Br(J/\Psi \rightarrow \mu^+ e^-) <8.3 \times 10^{-6}.
\end{eqnarray}
  
Recently Nussinov, Peccei and I[4] have examined  ``unitarity inspired"
relations between two- and
three-body lepton flavor violating decays and found that 
 the existing strong bounds on 
$\mu \to 3e$ and $ \mu\to e\gamma\gamma$ severly 
 constrain two-body lepton flavor violating decays of vector bosons
[$J/\psi,~\Upsilon$, and
$Z^o$] or pseudoscalars [$\pi^o,\eta$] into $\mu^\pm e^\mp$ final states.
However
the
bounds derived in Ref.[4] can be avoided if there is a kinematical
suppression or as a result of some
cancellations.
 Searching
for lepton flavor violating decays of vector bosons such as $J/\Psi$ 
remains a worthwhile experimental challenge.

BES started one year ago[5] and has been working on an experiment
of searching for $J/\Psi \rightarrow e^\mp \mu^\pm$. They will publish
their result officially in the near future.

\section{\bf {Single D meson production in $J/\Psi$ decays}}
Kinematically $J/\Psi$ can not decay into D meson pairs, however it is
able
to decay into single D meson. In the standard model, these Cabbibo
suppressed and/or favored weak decays have a typical
branch ratio $\sim 10^{-8}$ or smaller, which is unobservable and because
of which these processes serve as a probe of new physics. Recently Datta,
O'Donnell, Pakvasa and I[6] have studied the possibility of searching for
new flavor changing neutral current in the decay of $J/\Psi$. 
The purpose of the study is to answer whether new physics can enhance
sufficiently
for the processes to be observable in the near future experiments. We
first perform a model independent analysis, then examine the
predictions of the models, such as TopColor models, minimal sypersymmetric
standard model with R-parity violation and the two Higgs doublet model. We
found that the branch ratio of $J/\Psi \rightarrow D/ {\overline D} X_u$
could be as large as $10^{-5}$[6].
  
Experimentally with BES-II data $\sim 8 \times 10^{6} J/\Psi$, BES found
no signal of single D production in $J/\Psi$ decays, and put limits on
decay rates [7](at $90\%$ C.L. ): $Br ( J/\Psi \rightarrow {\overline D^0}
\rho^0
) \leq 1.25 \times 10^{-5}; ~~~~~ Br (J/\Psi \rightarrow D_s^+ K^- )
\leq 2.8 \times 10^{-5}$, and  $Br( J/\Psi \rightarrow D / {\overline D}
X_u ) \leq   
1.7 \times 10^{-4}$. These are preliminary results.

Next year BES will have collected $\sim 5 \times 10^7 J/\Psi$. The upper
limit on $Br( J/\Psi \rightarrow D / {\overline D} X_u )$
is expected to be reduced to $\sim 3 \times 10^{-5}$, which is very close
to the theoretical prediction. If there is still no signal found, it
will put some constraints on models beyond the standard electroweak
theory[6]. 

\section {\bf {CP violation in $J/\Psi$ decays }}

The origin of CP violation remains one of the outstanding problems in
particle physics and cosmology. To pin down the sources and nature of CP
violation in or beyond the Cabbibo-Kabayashi-Maskawa model, it would be
necessary to consider different observations of CP violation in different
channels from the K system, B sysytem, {\it etc}. The reaction of
interest at BES is[8] 
$$ e^+(p) + e^-(-p) \rightarrow J/\Psi \rightarrow A(q_-) + {\bar B(q_+)} 
+ X 
~~~~~, ~~~ (5)$$
where A (${\bar B}$) are charged particles, for instance $\pi^\pm$ in the
processes
of three or five pions decay of the $J/\Psi$, and $p, q_- $
and $q_+$ are
corresponding momentum in the laboratory frame. Define CP/T-odd operator
  $$ O_1 ~~ = ~~\frac{{\vec p} \cdot {\vec q_+} \times {\vec q_-}}{
    | {\vec p} \cdot {\vec q_+} \times {\vec q_-} | }~~. ~~~ (6) $$
If there exists CP violating interaction in $J/\Psi$ decays, one would
expect a non-vanishing expectation value of operator $O_1$.   
Theoretically there could be many sources responsible for the CP
violation in the process. One of them is the
Chromo-dipole moment of the charm quark
  $ i g_c \frac{d_g}{2} {\overline c}\sigma_{\mu\nu}
\gamma_5 {\lambda^a} c G_a^{\mu\nu} $ where $d_g$ has ${\rm [mass]}^{-1}$ 
dimension and
the $g_c$ the strong interaction coupling constant. 

Consider the process of three pion decay of the $J/\Psi$. Its
branch ratio is around $1.5\%$. With $10^8 \sim 10^9 J/\Psi$ at BEPC
II there will be around $10^6 \sim 10^7$
available
for the analysis of CP violation. To estimate the
experimental sensitivities to $d_g$, let us consider only the statistical
uncertainties. 
Neglecting the systematical uncertainties one expects to be able at BES to
probe for $d_g$ as
small as $ ( 1/{\sqrt{10^6 \sim 10^7}}) \times ({1 / m_c}) \leq 10^{-17}
{\it  cm}$.

One can easily construct different kind of  
CP/T-odd operators of three momentum products ($p, q_-, q_+ $) for the
analysis of CP
violation in $J/\Psi$ decays. For instance,
$$ O_2^{i j} \sim {( {\vec q_+ } - {\vec q_-} )}^i \cdot 
               {( {\vec q_+} \times {\vec q_-} )}^j + (i <-> j). ~~~~
(7)$$

The initial electron and/or positron beams are not polarized at BEPC,
otherwise one would be able to construct observables with the initial
polarization vector $\vec \sigma$, 
  $$ O_3 \sim {\vec \sigma} \cdot {\vec q_-} \times {\vec q_+}.~~~~~~~~~~~ 
(8) $$

 There has been proposal to measure CP violation in $e^+ e^-
\rightarrow J/\Psi \rightarrow {\Lambda} {\bar\Lambda}$. With a large
sample of $J/\Psi$, one expects to probe for and put a strong bound on the   
electric dipole moment of $\Lambda$[9].

\section{Conclusion and comments}

In this paper I have concentrated on three kind of processes to probe for
new
physics in the decay of $J/\Psi$. There are some other rare decay modes
which are interesting, but not reviewed. For instance, 
with a large sample of $J/\Psi$, it is possible and physically
interesting to search for Goldstone or Pseudo-Goldstone J, such as Axion
in the process 
$$ J/\Psi \rightarrow \gamma + J ~~~~~~.  ~~~~~~~~~~(9)$$ Another example
is
the invisible decay of
$J/\Psi$ investigated by Chang, Lebedev and Ng[10] recently 
in models with extra Z-bosons, minimal sypersymmetric
standard model with R-parity violation and decays into Goldstinos. The
third example is provided by Bijnen and Maul[11], who 
 recently have calculated in
detail the branch ratio of $J/\Psi$ decay into photon + missing energy
in the
popular theory these days
with large extra dimension. They 
found that the branch ratio could be as large as $10^{-5}$, which is
measurable at
BES. Before conclusion, we point out that even though we focus our
discussions here on $J/\Psi$
decays,
it is quite easy to apply the studies in this paper for $\Psi^\prime$ and
$\Upsilon$ system[12].

\begin{center} {\bf ACKNOWLEDGMENTS} \end{center}
I am grateful to my collaborators and colleagues 
for discussions. This work is
supportted in part by the NSF of China.

\begin{center}{\bf REFERENCES} \end{center}

\begin{enumerate}

\item Particle Data Group, C. Caso ${\it et al.}$, Europ. Phys. J, C3, 1
(1998).

\item "Probing for lepton flavor violation in decays of charmonium
and
bottomonium systems", R.D. Peccei, Jian-Xiong Wang and Xinmin Zhang, May
1998 Note (unpublished); Xinmin Zhang, invited talked given at the
national conference on high energy physics, Chengde, China, April (1998).

\item For examples, see, Z.K. Silagadze, hep-ph/9907328, July (1999);
T. Huang, Z. Lin and X. Zhang, hep-ph/0009353 (2000); Chan Hong-Mo et al,
hep-ph/0006338, hep-ph/0007004, hep-ph/0008313/0008324.

\item S. Nussinov, R.D. Peccei and X. Zhang, Hep-ph/0004153, April
(2000).

\item G. Tong et al (Private communication).

\item A. Datta, P.J. O'Donnell, S. Pakvasa and X. Zhang, Phys. Rev.
D60,
014011 (1999).

\item G. Rong et al (in
preparation).

\item Xinmin Zhang, Jian-Xiong Wang, Jian-Pin Ma, Dongshen Du and
Wu-Jun Huo
(in preparation).

\item Xiao-Gang He, Jian-Pin Ma and McKellar, Phys. Rev. D49, 4548
(1994); Ye
Yiun-Xiou and Ye Zheng-Yu (unpublished).

\item N.
Chang, O. Lebedev and J.N. Ng, hep-ph/9806497, June (1998). 

\item J. Bijnens and M. Maul, hep-ph/0006042, July (2000).

\item For example, it has been proposed to probe for CP violation
in the
process $\Psi^\prime \rightarrow J/\Psi + \pi^+ + \pi^- $, [ Xinmin Zhang,
Dongshen Du, Pin Wang, A. Datta, Jian-Xiong Wang and Jian-Pin Ma, May 1998
note, unpublished; The experimental analysis started already(Jin Li and
Zhi-Jin Guo, private communication)].

\end{enumerate}

\end{document}